\documentclass[useAMS,usenatbib]{mn2e}
\usepackage{color}
\usepackage{graphicx}
\usepackage{multirow}

\newcommand{\apj}{ApJ}
\newcommand{\apjl}{ApJL}
\newcommand{\mnras}{MNRAS}
\newcommand{\aap}{A\&A}

\newcommand{\nat}{Nature}
\newcommand{\pasp}{PASP}

\title{An inhomogeneous jet model for the rapid variability of TeV blazars}
\author[T.~Boutelier et al.]
{T.~Boutelier$^1$, 
G.~Henri$^1$, 
P--O.~Petrucci$^1$ 
\\ $^1$Laboratoire d'Astrophysique de Grenoble--Universit\'eŽ Joseph-Fourier/CNRS UMR 5571 --BP~53, F-38041 Grenoble, France}
\date{Accepted 2008 July 29}
\pagerange{\pageref{firstpage}--\pageref{lastpage}} \pubyear{2008}

\def\BLF{bulk Lorentz factor }
\def\BDF{Doppler factor }

\def\pks{PKS~2155--304 }

\begin{document}
\label{firstpage}

\maketitle

\begin{abstract}
We present a new time-dependent inhomogeneous jet model of non-thermal blazar emission, which reproduces the entire spectral energy distribution together with the rapid gamma-ray variability. Ultra-relativistic leptons are injected at the base of a jet and propagate along the jet structure. We assume continuous reacceleration and cooling, producing a relativistic quasi-maxwellian (or "pile-up") particle energy distribution. The synchrotron and Synchrotron-Self Compton jet emissivity are computed at each altitude. Klein-Nishina effects as well as intrinsic gamma-gamma absorption are included in the computation. Due to the pair production optical depth, considerable particle density enhancement can occur, particularly during flaring states.Time-dependent jet emission can be computed by varying the particle injection, but due to the sensitivity of pair production process, only small variations of the injected density are required during the flares. 
The stratification of the jet emission, together with a pile-up distribution, allows significantly lower bulk Lorentz factors, compared to one-zone models. Applying this model to the case of \pks and its big TeV flare observed in 2006, we can reproduce {\it simultaneously} the average broad band spectrum of this source as well as the TeV spectra and TeV light curve of the flare  with  \BLF lower than 15.

\end{abstract}

\section{Introduction}\label{intro}
It is widely admitted that the blazar phenomenon is due to relativistic Doppler boosting of the non-thermal jet emission taking place in radio-loud Active Galactic Nuclei (AGN) whose jet axis is closely aligned with the observer's line of sight. { Blazars} exhibit very broad spectral energy distributions (SED) ranging from the radio to the gamma-ray band { and dominated by two broad band components}. In the Synchrotron Self Compton scenario (SSC), the lowest energy hump is attributed to the synchrotron emission of relativistic leptonic particles, and the highest one is attributed to the Inverse Compton process (IC) of the same leptons on the synchrotron photon field. Broad band observations of these objects are crucial to understand the jet physics and to put reliable constraints on jet parameters.

The most extreme class of blazars are the highly peaked BL lac sources (HBL), where the synchrotron/Inverse Compton components peak in the UV/X-ray/gamma-ray (GeV up to TeV)  range.   The recent development of observational techniques in the TeV range, with Cherenkov telecopes experiments like HESS, MAGIC or VERITAS  has allowed the detection of about 18 HBL above 300~GeV. These objects are well known to be highly variable in all energy bands, from radio to gamma-ray, with timescales varying with energy. Perhaps the most extreme and remarkable example of this extraordinary variability behaviour has been caught by the HESS instrument in the recent observations of \pks during summer 2006 \citep{2155Flare} (impressive observations has also been recently obtained for Mkn 501 i.e. \citealt{Mrk501Flare}).

The most simple models of high energy emission assume a one-zone, homogeneous region. The SSC emission is assumed to come from  a spherical zone of radius $R$, filled with relativistic leptons characterized by their density $N$ and a characteristic Lorentz factor $\gamma$ (e.g. that most contributing to the peak emission), embedded in a isotropic magnetic field $B$. The blob is assume to move with a \BLF ${\Gamma_b}$  yielding to a \BDF $\displaystyle\delta_b=[\Gamma_b(1-\beta\cos\theta)]^{-1}$  where $\theta$ is the angle between the blob direction of motion and the line of sight. In this model, all physical quantities are averaged on the whole spherical region, and, more importantly, the synchrotron and IC emission are cospatial.   Causality constrains and a necessarily low pair creation optical depth then generally implies relatively high  bulk Lorentz factor \citep{Mastic97}.

In the case of the 2006 big flare of \pks \citep{2155Flare},  the observed variability time scale ($\sim$ 200 sec) in the TeV range implies a minimum \BLF greater than 50 \citep{Beg07} assuming an homogeneous one zone model. However, such high values of the \BLF are in contradiction with constrains derived from other observational evidence (\citealt{Pad95,HS06} and references therein). Furthermore, one-zone models are unable to fit the entire spectrum, the low energy radio points being generally attributed to more distant emitting regions.  
More complex models have been proposed including for example jet stratification (\citealt{Ghi85,Katar03}) or jet deceleration (e.g. \citealt{GeorgKaz03}).
We present here {a new} approach, unifying small and large scales emission region:  we consider that the radio jet is actually filled by the same particles originating from the high energy emitting region, at the bottom of the jet, that have propagated along it. We describe thus the emitting plasma by a continuous (although variable) particle injection, submitted to continuous reacceleration and radiative cooling. This model fits well into the two-flow framework originally proposed by \citet{Pel85} and \citet{Sol89} (see also \citealt{Tsin02} or the "spine-in-jet" model developped by \citealt{Chia00}) where a non relativistic, but powerful MHD jet launched by the accretion disk, surrounds a highly relativistic plasma of electron-positron pairs propagating along its axis. The MHD jet plays the role of a collimater and an energy reservoir for the pair plasma, which is responsible for the observed broad band emission. The present model only concentrates on the physical parameters of the relativistic pair beam and is described in Sect.~\ref{sec:model}. We show its application to the case of \pks in Sect.~\ref{sec:apli} focusing on the 2006 big flare event. We then discuss our results in Sect.~\ref{sec:disc}.

\section{Description of the model}
\label{sec:model}
\subsection{Geometry of the model}
We consider  that the relativistic plasma propagates in a stationary funnel whose geometry is parametrized as follows:
\begin{equation} \label{eq:R}
r(z)=R_0\left[\frac{z}{Z_0}+\left( \frac{R_i}{R_0} \right)^{1/\omega} \right]^{\omega}
\end{equation}
where $r(z)$ is the radius of the jet at the altitude $z$. This shape describes a jet with a "shifted" paraboloid shape, with an initial inner radius $R_i$ at $z=0$,   and a radius $R_0$ at a distance  $Z_0$ from the apex.  The index $\omega$ is lower than 1 for a collimated jet. The magnetic field inside the jet is radially averaged at each altitude, and has the following dependence:
\begin{equation} \label{eq:B}
B(z)=B_0\left(\frac{r(z)}{R_0}\right)^{-\lambda}
\end{equation}
We take  $\lambda$ between $1$ and $2$, the two extreme values describing respectively a pure toroidal or a purely poloidal field distribution.

We consider that the jet is continuously accelerating, starting from rest ($\Gamma_b = 1$ at $z=0$) and reaching an asymptotic value $\Gamma_{b \infty}$. The acceleration is assumed to take place over a distance comparable with $Z_0$. A detailed model of the acceleration mechanism is beyond the scope of this paper, so we chose a simple parametrization to describe the evolution of $\Gamma_b(z)$:
\begin{equation} \label{eq:Gam}
\Gamma_b (z)=\left[ 1+\frac{\Gamma_{b \infty}^a -1}{1+\frac{z_0}{z}}\right]^{1/a}
\end{equation}
$a$ being a "stiffness" parameter describing the width of the accelerating region. 
\begin{figure}
\includegraphics [angle=90,width=84mm]{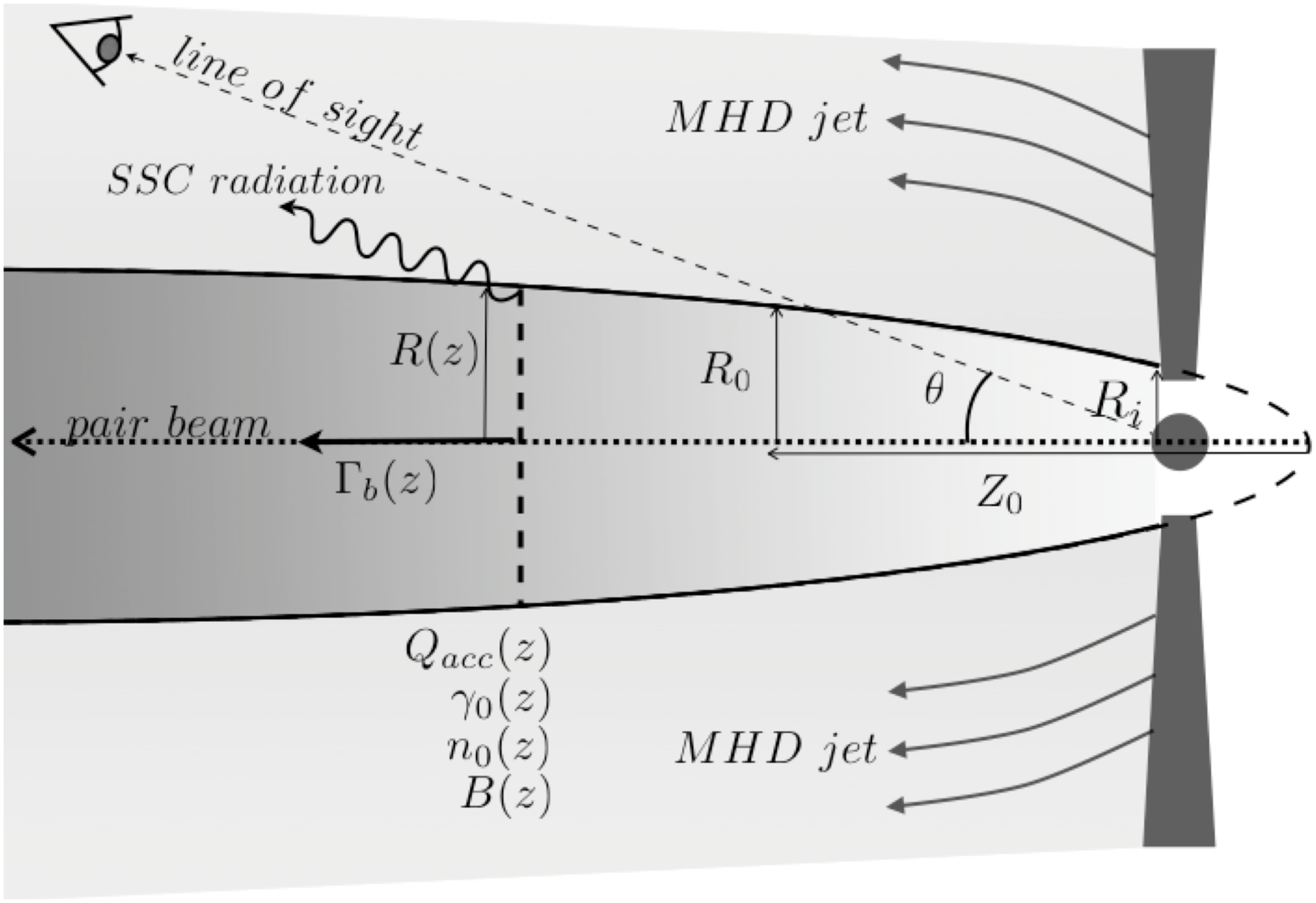}
\caption{Sketch of the jet geometry. See text for the signification of the different parameters.}
\label{fig:geo}
\end{figure}

\begin{table*}
\caption{Model parameters of the flaring and quiescent state. During the flaring state, the flux of injected particles varies between the indicated minimum and maximum values, following the injection pattern displayed in Fig. 3. The other parameters remain fixed. $R_{i}$, $R_{0}$, $Z_{0}$, $Z_{c}$ are in unit of $10^{14}cm$}
\label{tab:param}
\vspace*{-0.5cm}
\begin{center}
\begin{tabular}{cccccccccccccccc}
\hline
STATE && $\Phi(Z_i)$ & $N(Z_{i})$ &$\Phi(Z_0)$ & $N(Z_{0})$ & $Q_0$ &$\Gamma_{b \infty} $ & $R_i$&$R_0$ & $Z_0$ & $Z_c$ & $B_0$ & $\omega$ & $\lambda$& $\zeta$  \\
  & & $[10^{42} s^{-1}]$& $[cm^{-3}] $ &$[10^{42} s^{-1}]$ & $[cm^{-3}]$&$[s^{-1}]$& & $$ &  &  &  & $[G]$ &  &  &    \\  
\hline
\multirow{3}*{ flaring} & Max&$2.09 $& $1817$ & $70.1 $& $1300$ & \multirow{3}*{ $6.5$}\\ 
 & Aver.&$1.84 $& $1666$ & $24.4$ & $600$ &\\ 
  & Min&$1.16 $& $1055$ &$2.33 $& $60$ & \\  \cline{1-7}
{ quiescent} & &$1.16$ & $1051$ &$1.55$& $40$ & $2.5$ & \multirow{-2}*{15} & \multirow{-2}*{$1.1$}& \multirow{-2}*{$1.78$}&\multirow{-2}*{$20$}& \multirow{-2}*{$5\times 10^{7}$}&\multirow{-2}*{5}&\multirow{-2}*{0.2}&\multirow{-2}*{1.9}&\multirow{-2}*{1.27} \\  
\hline
\end{tabular}
\end{center}
\end{table*}

\subsection{Particle energy distribution}
In our model, the energy distribution function (EDF) of the electron-positron plasma is assumed to be a relativistic maxwellian (or "pile-up") distribution :
\begin{equation} \label{eq:N}
n(\gamma,z,t)=n_0(z,t)\gamma^2\exp\left(-\frac{\gamma}{\gamma_0(z,t)}\right)
\end{equation}
where   $n_0(z,t)$ is a normalization factor and $\gamma_0$ the characteristic "pile-up" Lorentz factor. This distribution is a natural outcome of some acceleration processes like second order Fermi acceleration or magnetic reconnection (\citealt{hen91}; \citealt{Sch84,Sch85}; \citealt{SH04}) . It has been also shown by \citet{SH04} that a pile up distribution is well suited to reproduce the narrow peaked high energy component of TeV blazars. 

When the plasma propagates in the structure, the particles loose energy via synchrotron and Inverse Compton cooling, producing the observed emission. It turns out however that the cooling time is much too short for the relativistic particles to fill the whole jet. It is thus necessary to assume a fast reacceleration process along the jet. The acceleration rate is parametrized by a shifted power law with an index $\zeta$ and an exponential cut-off after some altitude $Z_c$ to avoid energy divergence:
\begin{equation} \label{eq:Q}
Q_{acc}(z)=Q_0\left[\frac{z}{Z_0}+\left( \frac{R_i}{R_0} \right)^{1/\omega}\right]^{-\zeta}\exp\left(-\frac{z}{Z_c}\right)
\end{equation}
In the two-flow framework (see introduction), this re-heating is naturally provided by the surrounding MHD structure via second order  Fermi process. Then we only need to determine two parameters to fully describe the relativistic plasma: $n_0(z,t)$ and $\gamma_0(z,t)$; $\gamma_0(z,t)$ is determined by balancing radiative losses and re-acceleration, and $n_0(z)$ by a continuity equation as explained in the next paragraph.

\subsection{Pair production}
We compute at each altitude and frequency the pair production optical depth in the comoving frame. Absorption of $\gamma$-ray photons induces the formation of new pairs, that are supposed to be continuously reaccelerated  as explained above. It results then in an increase of the particle flux $ \displaystyle{\Phi (z,t) = \int n(\gamma,z,t) S(z) \Gamma_b \beta_b c d\gamma}$, which is conserved in the absence of pair creation. $\Phi(z)$ is computed through a continuity equation : the importance of pair creation is measured by the variation of this flux. The pair production optical depth is smaller than that expected in one-zone models, where soft photons are produced cospatially  with higher energy ones. Indeed a relativistic Maxwellian particle distribution results in a much lower local soft photon density, compared to a power law spectrum :  soft photons are produced farther away in the jet and do not contribute to opacity. Consequently it allows  significantly lower bulk Lorentz factor compared to the ones obtained with simple one-zone models. In the case of \pks our results are compatible with an asymptotic \BLF of about 15, much below the value of 50 inferred by some authors (e.g. \citealt{Beg07}). It turns out that pair production plays a  fundamental role in explaining the flares in gamma-rays, because when the optical depth is close to one, a very small variation of the initial particle density can trigger a considerable enhancement in the pair production region.

 \subsection{The jet spectrum}
Knowing $n_0(z,t)$, $\gamma_0(z,t)$, $r(z,t)$ and $B(z,t)$, we compute the emissivity at each altitude in the jet by assuming a SSC process for the radiative mechanism. The total intensity of the jet is then determined by integrating the emissivity all along the jet.
The emissivity is enhanced in the observer frame by the Doppler boosting. We take also into account the attenuation of the gamma ray signal by the cosmic diffuse infrared background, chosen as the "modified" Primack model P45 \citep{EBL06}.

Once injected at the base of  the jet, the particles contribute first to the high energy part of the jet SED. As they propagate, their emissivity peaks progressively  at lower energy, producing the low energy part of the spectrum. Hence, the jet can be seen as a continuous succession of time dependent one-zone SSC models that propagate inside a stationary geometry, each of them contributing with its particular spectrum to the whole observed SED. The spectral shape of the whole SED is not controlled by the local particle energy distribution (which is always locally a narrow pile-up), but rather by the z-dependencies of the jet radius, the magnetic field, and the acceleration rate. 
A constant injection rate would lead to a stationary emission pattern,  which would be rather easy to fit. In reality however, the observed instantaneous spectra are a complicated convolution of the whole history of the jet, keeping the memory of the whole past (and unknown)  injection pattern. We describe below a simple procedure to extract the physical parameters of the jet from observed spectra, despite the fact that they do not correspond to a simple steady-state of the jet.

\subsection{Time dependent simulations. }
Real observations result from a complex combination of the whole injection pattern at the basis of the jet. 
Due to the propagation of particles, the high energy part of the synchrotron and TeV components will be dominated by a single flare, occuring at the basis, whereas  the low energy part of the spectrum is a convolution over a large scale of the past jet history: it is thus rather {a time-averaged} spectral state mixing quiescent and flaring states. The instantaneous spectrum is thus a mix of different ideal "steady-states". 

To constrain the geometrical parameters, we first build a virtual set of data called the fake flaring spectrum, that would be observed if the jet were constantly flaring. 
 The high energy part of this spectrum is the really observed emission during a flare. On the other side, its low energy part corresponds to the  observed data corrected by an enhancement factor to take into account that the actual jet is flaring only a fraction $f$ (called duty cycle) of the time. This enhancement factor is then $f^{-1}$ since the real jet is filled
 only partially with the high particle density associated to a flare, whereas the fake spectrum corresponds to a fully filled jet.
For the intermediate (X-rays) energy bands, the procedure is not so accurate because a small (but statistically variable) number of flares can contribute to the flux. 
We chose to adjust approximately the spectra by some intermediate factor in this band, but we have checked that our results are not very sensitive to this points. 
The $f$ factor can be determined observationally from the fraction of time we see the object flaring within statistically random observation periods. 
Once the "fake flaring" spectrum is constructed, we use it to adjust the jet model geometrical parameters, and the flaring injection rate. We take also into account the required variability timescale which constrains basically the jet size and the Doppler factor.
The variability timescale is commonly evaluated in one-zone models by the light travel time of the zone divided by the Doppler factor $\displaystyle R/\delta c$. However, in the case of a stationary structure like a jet, the correct estimate is to take the typical length of the emission zone , corrected for the time contraction due to the difference in light travel time to the observer along different parts of the jet, like for the superluminal motion. This gives $\displaystyle t_{var} \simeq \frac {Z_0}{c}(1-\beta_b \cos \theta) = \frac {Z_0} {c \Gamma_b \delta} $ which is of the same order of the one-zone (transverse) variability timescale if we assume $R_0/Z_0 \simeq \Gamma_b^{-1}$. The model predicts also naturally an increasing variability with energy since more and more individual flares contribute to lower energy ranges. This is indeed observed in TeV blazars \citep{GD07}. 
Once the flaring state parameters are found, we construct a quiescent state by keeping the same jet geometry, but reducing the particle injection rate and/or acceleration rate to fit low flux observations. A general light curve can then be obtained by adjusting an injection pattern oscillating around the average flaring state, for a given particular flare.

\vspace*{-0.5cm}
\section{Application to \pks}
\label{sec:apli}

We have tested our model to  the big flare event observed by HESS in July 2006 in \pks \citep{2155Flare}, the strongest flaring event ever seen in the TeV range for a blazar. During this flare  the average flux  above 200~GeV of the source reached about $\sim~7$ crab. Moreover variability with timescales as small as $\sim$200 seconds have been observed. The light curve is very well sampled (see Fig.~\ref{fig:LightCurve}), as well as the spectrum, which makes this event particulary interesting to test models.

We apply the method described previously to the SED of PKS~2155--304.  We have first constructed an average spectrum of \pks in the low frequencies range ($\la10^{15}~Hz$), by compiling available archival data (from the HEASARC archive website (http://heasarc.gsfc.nasa.gov/docs/archive.html)). Based on the activity detected by HESS, we estimate that the duty-cycle is around 10\%. An accurate value is not necessary since it would modify only the fake spectrum and not the real one.
Then to construct the "fake flaring" spectrum of this source we combine the HESS data with the { radio-to-optical} ones corrected by a factor 10.
To better constrain the space parameter we include also archival X-ray data 
from BeppoSAX, XMM-Newton and SWIFT at different epochs \citep{Mas07}, as well as archival EGRET data\citep{Ver95}. We choose not to re-normalize these data by any intermediate "duty-cycle" factor. Then  they would correspond { respectively to lower limits of the "flaring state" X-ray and GeV flux}.

\begin{figure}
\begin{center}
\includegraphics[angle=0,width=0.9\columnwidth]{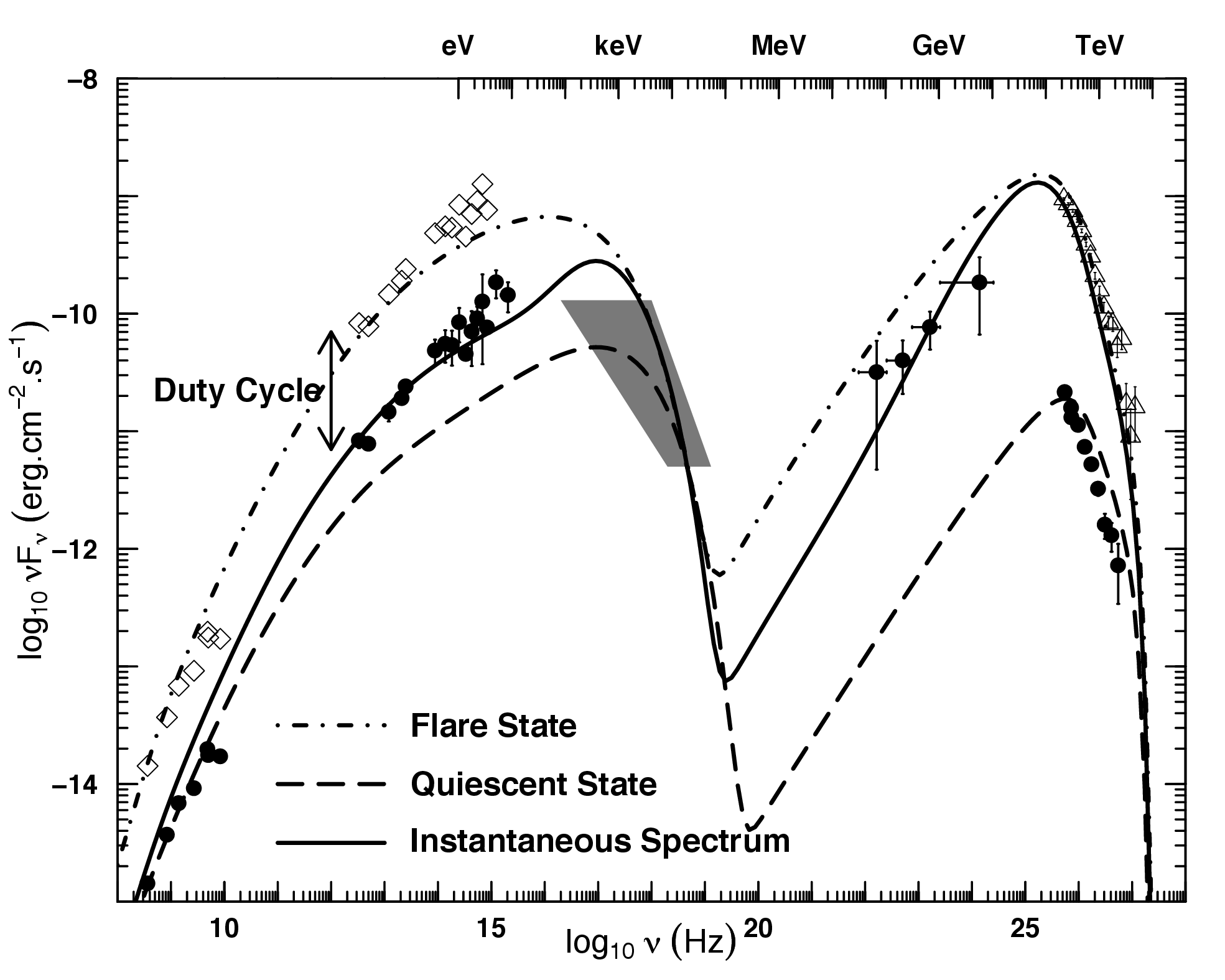}
\end{center}
\caption{Fit of \pks data. {\bf Filled dots:} average archival data (see text). {\bf Empty triangles:} average HESS data from the big flaring night. {\bf Empty diamonds:} "fake flaring" state low energy points  with a duty cycle f=0.1 (see text) {{\bf Shaded area:} enveloppe of archival X-ray data from BeppoSAX, SWIFT and XMM-Newton}. {\bf Dot-dashed line:} best fit of the " average fake flaring" spectrum. {\bf Dashed line:} fit of the { quiescent ($\sim$average)} spectrum. {\bf Solid line:} example of an instantaneous  {\bf simulated spectrum}.}
\label{fig:fit}
\vspace*{0cm}
\end{figure}

An average "fake flaring" spectrum is shown in Fig.~\ref{fig:fit} { in dot-dashed line}. The corresponding best fit model parameters are reported in Tab. \ref{tab:param}. Interestingly we only needs a \BDF $\Gamma_b=15$ which is significantly below the values of $\sim$ 50 inferred from { one-zone model} \citep{Beg07}.  For this simulation, the characteristic time scale ${R_0/(\delta_bc)}$ { is $\sim$200} seconds in agreement with observations.

Once we have obtained the best fit parameters of the "flaring state", we fit the TeV "quiescent state" by fixing the fit parameters associated with the jet structure and dynamics, allowing only the density of the particles filling the jet free to vary. We choose to fit the averaged spectrum derived above 200~GeV by HESS in \citet{2155_05}, since it does not strongly differ from the quiescent state. The corresponding best fit parameters are also reported in Tab.~\ref{tab:param} and the best fit model has been overplotted  in Fig.~\ref{fig:fit} { in dashed line}.  

\subsection{Time dependent simulation}
\begin{figure}
\begin{center}
\includegraphics[width=0.9\columnwidth]{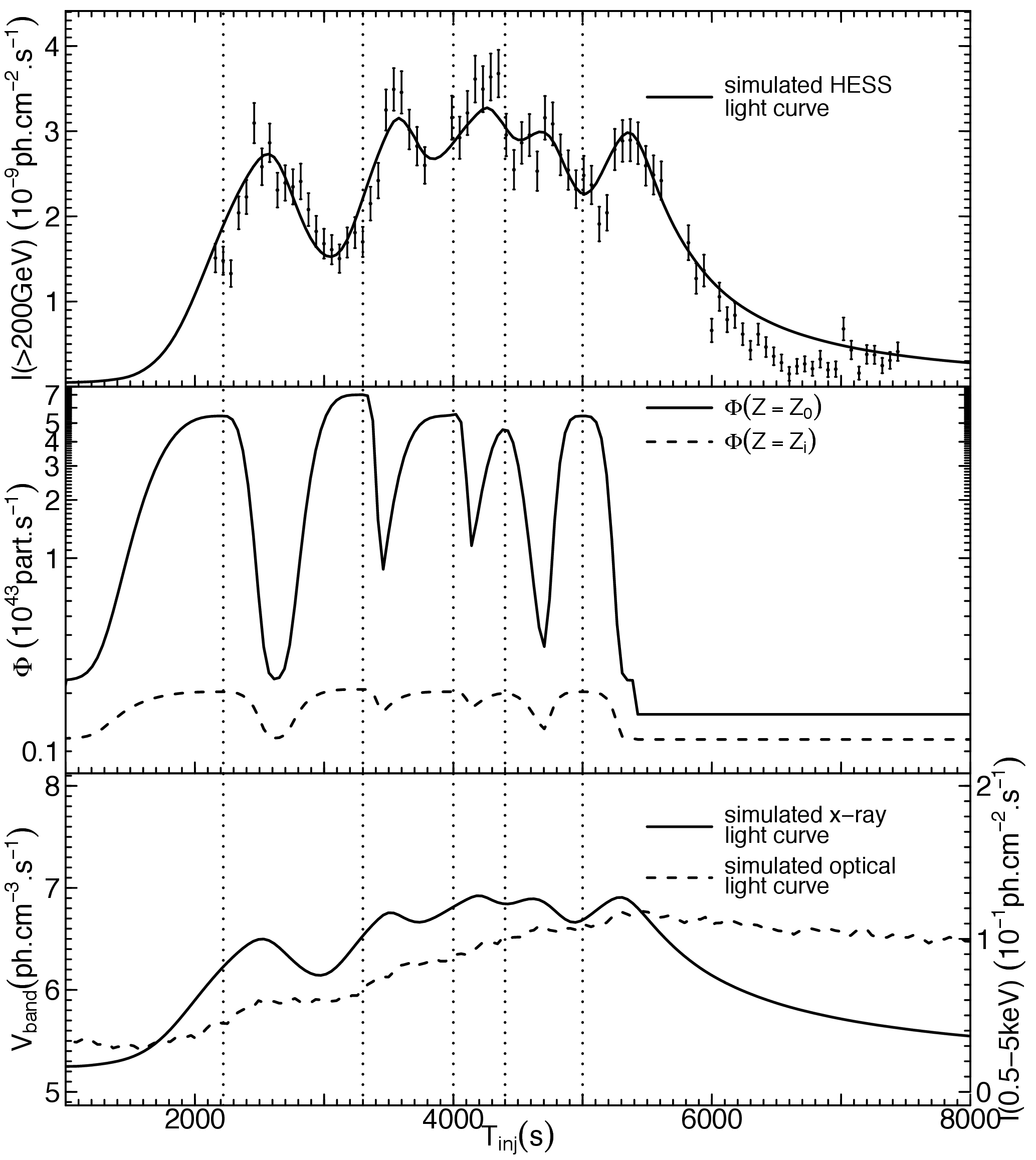}
\end{center}
\vspace*{-0.5cm}
\caption{{\bf Upper panel}: HESS light curve above 200~GeV superimposed with the model (solid line). {\bf Middle panel}: time dependent particle injection function used in the simulation. {\bf Lower panel}: predicted light curves in X-ray (dashed line, left y-scale) and optical (dot-dashed line, right y-scale). The dotted lines mark the maximum of the different bursts of the injection function.\vspace*{-0.5cm}
}
\label{fig:LightCurve}
\end{figure}

After having constrained the jet and plasma characteristics of \pks in both quiescent and flaring states, we aim at reproducing the lightcurve observed by HESS during the big flare event by using a variable injection of particle.  Based on the analysis of \citet{2155Flare}, where the light curve is decomposed in 5 successive assymetric bursts, we use, for the flare period, an injection function $\Phi(z=0,t)$ that is the sum of five "generalized Gaussian" shape \citep{Nor96}   e.g.:
\begin{equation} \label{eq:Ninj}
\Phi(0,t)= \sum_i \Phi_0^i\exp-\left(\frac{\left|t-t_{max}^i\right|}{\sigma_{r,d}^i}\right)^{\kappa^i}
\end{equation}
where $t_{max}^i$ is the time of the maximum of the burst $i$, $\sigma_{r}^i$ and $\sigma_{d}^i$ the rise $(t<t_{max}^i)$ and decay $(t>t_{max}^i)$ time constant respectively, and $\kappa^i$ is a measure of the sharpness of the burst. We have plotted { in the middle} of Fig. \ref{fig:LightCurve} the assumed injection function. Before the flare, $\Phi(z=0,t)$is assumed to be a crenel function that oscillates between quiescent and flaring states in agreement with the source duty cycle.
At the top of  Fig. \ref{fig:LightCurve} we have reported the HESS light curve and the simulated one. The agreement is very good. We have also overplotted in Fig. \ref{fig:fit} in solid line an instantaneous spectrum extracted from the time-dependent simulation { during the flare period}. It agrees nicely with the broad-band (from radio to TeV) spectrum observed { during this burst}.

\section{Discussion}
\label{sec:disc}

Our time-dependent inhomogeneous jet model succeeds in  reproducing {\it simultaneously} the broad band (from radio to TeV) spectrum of \pks as well as the TeV light curve during the big flare event of July 2006. The key idea of the method is to decompose the blazar spectrum in "quiescent", low luminosity states, and "flaring", high luminosity states. The high energy part of the spectrum, coming from  small-scale inner regions, is assumed to be, at any time, in one of these pure states. On the other hand the low energy part is a convolution over a large scale of the past history of the jet : it is thus rather a time-averaged spectral state mixing quiescent and flaring states in proportions given by the source "duty-cycle". Moreover we do not require two different populations of emitting particles like in other models (e.g. \citealt{Katar03}, \citealt{Chia00}) but simply a continuous (although variable) injection of particles at the base of the jet that propagate along the jet structure. Pair production plays an important role to amplify the initial variation, as can be seen with the variation of the particle flux along the jet during the flaring state (see Tab. 1) : the pair current is amplified by a factor 30 at the end of the jet, when the initial current varies only by a factor 2. 

The model can also predict light curves at different wavelengths. As an example, the X-ray (2-10 keV) and optical (V band) light curves expected during the TeV flare have been plotted at the bottom of Fig. \ref{fig:LightCurve}. The X-ray luminosity exhibits almost simultaneous variations but with a lower amplitude (about 5 times smaller). On the other hand, the optical light curve shows a very different behavior, increasing all along the flare. This is due to to the large size of the optical emitting region that plays the role of a low pass filter. Consequently, the optical luminosity integrates the recent past history of the jet.
These results are compatible with simultaneous multiwavelength observations made during the ``Chandra night'' (oral communication).

This model puts also some constraints on the central black hole mass. Indeed, the radius at the base of the jet is directly linked to the bulk Doppler factor. From our best fit parameters,  we find $M_{bh}= 8\times 10^8 {r_g/}{R_0}$, where $r_g=GM_{bh}/c^2$ is the gravitational radius. This implies a black hole mass  smaller than $10^9~M_{\odot}$ usually invoked in the case of \pks  \citep{Koti98}. We believe however that it is still acceptable given the mass measurement uncertainties \citep{Bet03}. {}

In the present work, we do not specify the origin of the variability. Obviously, other plasma parameters can vary in addition to the injection density $N_0$. Most likely, variability  can be triggered by a change in the acceleration rate described by $Q_0$ (Eq. \ref{eq:Q}). In a plausible scenario, long term (year scale) variability implying the succession of quiescent and active states could be attributed to variations in the accretion rate, wheras the short (minute-scale) flares would be attributed to the instability to pair creation that develops only when the initial particle density is close to a critical threshold.
\vspace*{-0.5cm}


\label{lastpage}
\end{document}